\documentclass[]{iopart}

\usepackage{color}
\usepackage{iopams}  
\usepackage{graphicx}

\begin{document}

\title{Breaking of four-fold lattice symmetry in a model for pnictide superconductors}

\author{M Daghofer and A Fischer}

\address{IFW Dresden, P.O. Box 27 01 16, D-01171 Dresden, Germany}
\ead{m.daghofer@ifw-dresden.de}
\begin{abstract}
We investigate the interplay of onsite Coulomb repulsion and various
mechanisms breaking the fourfold lattice symmetry in a three-band
model for the iron planes of iron-based superconductors. Using 
cluster-perturbation theory allows us to locally break the
symmetry between the $x$- and $y$-directions without imposing
long-range magnetic order. Previously investigated anisotropic
magnetic couplings are compared to 
an orbital-ordering field and anisotropic hoppings. We find that all
three mechanisms for a broken rotational symmetry lead to similar
signatures once onsite interactions are strong enough to bring the
system close to a spin-density wave. The band
  distortions near the Fermi level are independent of differences
  between the total densities found in $xz$ and $yz$ orbitals.
\end{abstract}

\pacs{74.70.Xa,74.25.Jb,71.27.+a,71.10.Fd}

\section{Introduction}\label{sec:intro}

Like cuprate superconductors, many of their iron-based cousins have an
antiferromagnetic (AFM) phase in the phase diagram near the superconducting
phase~\cite{pnict_rev_johnston,Paglione:2010p2493}. As phonons are
moreover not believed to be strong enough to explain the relatively
high transition temperatures of pnictides~\cite{phonon0}, the AFM
interactions have attracted large interest~\cite{chubukov_review}. Apart from the metallic
character of the spin-density--wave (SDW) phase in pnictides -- as
opposed to the Mott insulator in undoped cuprates -- a second
difference is that the SDW in pnictides breaks the four-fold
symmetry of the iron-arsenic planes down to a two-fold symmetry. 

This breaking of the four-fold lattice symmetry is seen in the
conductivity~\cite{rho_anisotropy,Anis_charge} and in angle-resolved
photo-emission spectroscopy
(ARPES)~\cite{Yi:PNAS2011,ARPES_NaFeAs11,He:2010pNaFeAs,ZhangARPES_NaFeAs_2011}
even at temperatures above the onset of magnetic
long-range order. While there is a structural phase transition at
slightly higher temperature and while the in-plane lattice constants
thus break the rotational lattice symmetry~\cite{neutrons1}, the
effects in experiments appear stronger than can be explained by
slightly different lattice constants. Additional symmetry-breaking of
the electronic degrees of freedom has thus been suggested to be
involved, especially (i) a breaking of the \emph{orbital} symmetry, i.e., a
lifting of the degeneracy of the $xz$ and $yz$
orbitals~\cite{kruger:054504,Lv10,oo_nematic_2011}, and (ii) a
\emph{nematic} phase of the spin degree of
freedom~\cite{Fernandes:2011transp,Fernandes_nematic2012}. In the
latter scenario, magnetic correlations already select the preferred ordering
vector, e.g., $(\pi,0)$ over the equivalent $(0,\pi)$, but do not yet
establish long-range magnetic
order~\cite{Fang:2008nematic,Xu2,Fernandes_nematic2012,PhysRevB.82.020408}. A
related picture involves clusters with short-range magnetic order
whose AFM direction is pinned by the lattice anisotropy~\cite{Shuhua_Pnict2011}.

One problem in deciding between the scenarios is precisely that they
all break the same symmetry, which implies that as soon as one of the
breaks rotational symmetry, the symmetry in the other degrees of
freedom will be broken as
well. Lattice~\cite{PhysRevB.79.180504,Paul:2011p2617,PhysRevB.82.020408}
and orbital~\cite{kruger:054504,Fernandes_nematic2012} degrees of
freedom thus strongly interact with the spin. Nevertheless,
identifying the signatures of various modes of symmetry breaking may
help to elucidate the most important effects. We calculate here the
spectral density of the three-orbital model, see
Sec.~\ref{sec:models}, where the four-fold rotational lattice symmetry
is broken by (i) orbital order, (ii) anisotropic hoppings and (iii)
short-range AFM interactions. In order to be able to keep the magnetic
interactions short-range and in order to include onsite Coulomb
interactions, we employ cluster-perturbation theory, extending an
earlier study~\cite{Akw_nematic} of anisotropy driven by short-range AFM and ferromagnetic
(FM) interactions. Anisotropic
  magnetic interactions can distort the spectral density in
agreement with ARPES both in otherwise noninteracting three- and
four-band models and for a regime close to the
SDW phase~\cite{Akw_nematic}, orbital order in the non-interacting model was discussed
in~\cite{oo_nematic_2011}. Here, we compare the two scenarios near the
SDW in more detail and also compare them to symmetry breaking via
anisotropic hoppings. While distortions induced by the latter cannot even qualitatively be
reconciled with ARPES for the noninteracting model, we are going to
see that results become more realistic near the SDW transition.

\section{Model and Method}\label{sec:models}

As we want to use exact diagonalization to solve a fully interacting
model on a small four-site cluster, we have to restrict the model to
at most three bands. We use a variant~\cite{Akw_nematic} of the
three-band model proposed in~\cite{Daghofer_3orb} that gives a
better fit of the Fermi surface. The unit cell of the Fe-As plane
contains two iron and two arsenic ions, however, an internal symmetry of
this unit cell allows us to write the Hamiltonian in terms of a
one-iron unit cell~\cite{plee,eschrig_tb} as long as we consider
isolated planes, as is done here. In momentum space, the tight-binding
Hamiltonian can then be written in terms of a pseudo-crystal
momentum ${\bf \tilde{k}}$, which is defined as ${\bf \tilde{k}} = {\bf k}$
for the $xz$/$yz$ orbitals and ${\bf \tilde{k}} = {\bf k}+(\pi,\pi)$ for the
$xy$ orbital and which is taken from the enlarged Brillouin zone
corresponding to a one-iron unit cell. When
  translating the spectral density $A({\bf k},\omega)$ 
from pseudo-crystal momentum ${\bf \tilde{k}}$ to `lab-space' momentum
${\bf k}$, spectral weight of $xy$ character is shifted by $(\pi,\pi)$
with respect to $xz$ and $yz$ weight~\cite{Tranquada_spinorb,oo_nematic_2011,Akw_nematic}.

The ${\bf \tilde{k}}$-dependent Hamiltonian in orbital space can then be written as 
\begin{eqnarray}\label{eq:H0k}
H_{\rm TB}(\mathbf{ \tilde{k}}) = \sum_{\mathbf{ \tilde{k}},\sigma,\mu,\nu} 
T^{\mu,\nu}(\mathbf{ \tilde{k}}) 
d^\dagger_{\mathbf{ \tilde{k}},\mu,\sigma} d^{\phantom{\dagger}}_{\mathbf{ \tilde{k}},\nu,\sigma}\;,
\end{eqnarray}
where $d^{\phantom{\dagger}}_{\mathbf{ \tilde{k}},\nu,\sigma}$
($d^{\dagger}_{\mathbf{ \tilde{k}},\nu,\sigma}$) annihilates (creates) an
electron with pseudo-crystal momentum ${\bf \tilde{k}}$ and spin
$\sigma$ in orbital $\nu$. The orbital indices $\nu,\mu=1,2,3$ refer to the $xz$,
$yz$ and $xy$ states of the iron $3d$ manifold, respectively. 
The $T^{\mu,\nu}(\mathbf{ \tilde{k}})=T^{\mu,\nu}(k_x,k_y)$ defining the hoppings are given
by 
\begin{eqnarray}\label{eq:ekin}
T^{11/22} &= 2t_{2/1}\cos  k_x +2t_{1/2}\cos  k_y \nonumber\\
 &\quad +4t_3 \cos  k_x \cos  k_y  \\
 &\quad \pm 2t_{11}(\cos 2k_x-\cos 2k_y)\nonumber\\
 &\quad+4t_{12}\cos 2 k_x \cos 2k_y,  \nonumber\\
T^{33} &= \Delta_{xy} + 2t_5(\cos  k_x+\cos  k_y)\\
&\quad +4t_6\cos  k_x\cos k_y  + 2t_9(\cos 2k_x+\cos 2k_y)\nonumber\\
&\quad + 4t_{10}(\cos 2k_x \cos k_y + \cos k_x \cos 2k_y),\nonumber\\
T^{12} &= T^{21} = 4t_4\sin  k_x \sin  k_y,  \label{eq:3b}\\
T^{13} &= \bar{T}^{31} = 2it_7\sin  k_x + 4it_8\sin  k_x \cos  k_y,  \\
T^{23} &= \bar{T}^{32} =  2it_7\sin  k_y + 4it_8\sin  k_y \cos  k_x,  
\end{eqnarray}
where a bar denotes the complex conjugate. Hopping parameters are the
same as in~\cite{Akw_nematic}: $t_1=-0.08$, $t_2=0.1825$,
$t_3=0.08375$, $t_4=-0.03$, $t_5=0.15$, $t_6=0.15$, $t_7=-0.12$,
$t_8=-t_7/2=0.06$, $t_{10} = -0.024$, $t_{11} = -0.01$, $t_{12} =
0.0275$, $\Delta_{xy} = 0.75$. The chemical potential $\mu$ depends on
the interaction terms and is chosen to fix the filling at 4 electrons
per site, for non-interacting electrons, we find $\mu= 0.47$. All
energies are given in eV. These bands are only an approximation and a
three-band model may not be detailed enough to capture
material-dependent properties.

We also include the onsite Coulomb interactions including Hund's rule
coupling and pair hopping~\cite{PhysRevB.18.4945,Oles:1983}. While the
couplings for the $xz$/$yz$ doublet can in 
principle differ from the ones involving the $xy$ orbital, we choose
them here to be the same and employ the symmetric relations $U=U'+2J$ for simplicity. 
\begin{eqnarray}  \label{eq:Hcoul}
  H_{\rm int}& =
  U\sum_{{\bf i},\alpha}n_{{\bf i},\alpha,\uparrow}n_{{\bf i},
    \alpha,\downarrow}
  +(U'-J/2)\sum_{{\bf i},
    \alpha < \beta}n_{{\bf i},\alpha}n_{{\bf i},\beta} \nonumber\\
  &\quad -2J\sum_{{\bf i},\alpha < \beta}{\bf S}_{\bf{i}\alpha}\cdot{\bf S}_{\bf{i}\beta}\\
&\quad +J\sum_{{\bf i},\alpha < \beta}(d^{\dagger}_{{\bf i},\alpha,\uparrow}
  d^{\dagger}_{{\bf i},\alpha,\downarrow}d^{\phantom{\dagger}}_{{\bf i},\beta,\downarrow}
  d^{\phantom{\dagger}}_{{\bf i},\beta,\uparrow}+h.c.), \nonumber
\end{eqnarray}
where $\alpha,\beta$ denote the orbital and ${\bf S}_{{\bf i},\alpha}$
($n_{{\bf i},\alpha}$) is the spin (electronic density) in orbital $\alpha$ at site
${\bf i}$. The electron-spin operators are given as usual by ${\bf S}_{{\bf i}\nu} = \frac{1}{2}\sum_{ss'}
d^{\dagger}_{{\bf i}\nu s}{\boldsymbol \sigma}^{\phantom{\dagger}}_{ss'}d^{\phantom{\dagger}}_{{\bf
    i}\nu s'}$, where ${\boldsymbol \sigma} = (\sigma^x,\sigma^y,\sigma^z)$
is the vector of Pauli matrices.
We choose here $U=1.02\;\textrm{eV}$ and $J=U/4$, because
the system is then very close to the SDW transition. The interacting
system has been found to be very susceptible to magnetic interactions
breaking rotational symmetry~\cite{Akw_nematic}, and we thus
concentrate on this regime.  

Similar to the approach chosen in~\cite{Akw_nematic}, we
explicitly break rotational symmetry by Heisenberg couplings that act
only locally within the small cluster directly solved with exact
diagonalization (see method below). Extending the analysis
in~\cite{Akw_nematic}, the coupling $J_{x}$ along the 
$x$-direction and $J_{y}$ along $y$ can have different magnitudes and
the same or different signs:
\begin{eqnarray}\label{eq:Heisenberg}
H_{\textrm{Heis}} &=  J_{x} \sum_{\stackrel{\langle {\bf i},{\bf
    j}\rangle\parallel x}{\mu\nu}}
{\bf S}_{{\bf i},\mu}\cdot
{\bf S}_{{\bf j},\nu} 
+ J_{y} \sum_{\stackrel{\langle {\bf i},{\bf
    j}\rangle\parallel y}{\mu\nu}}
{\bf S}_{{\bf i},\mu}\cdot
{\bf S}_{{\bf j},\nu},
\end{eqnarray}
where $\mu$, $\nu$ denote orbitals and $\langle {\bf i},{\bf
  j}\rangle\parallel x/y$ nearest-neighbour (NN) bonds along the $x$
and $y$ directions. For $J_{x/y}>0$, the coupling is AFM.  

We compare this magnetic symmetry breaking addressing the anisotropy
of the magnetic state to a an orbital symmetry breaking addressing the
symmetry between the $xz$ and $yz$ orbitals. The latter is implemented
as a difference in onsite energy
\begin{equation}\label{eq:oo}
H_{\textrm{orb}} = \Delta \sum_{\bf i}(n_{{\bf i},yz}-n_{{\bf i},xz}),
\end{equation}
where $\Delta>0$ favours occupation of the $xz$ orbital, as proposed as
an explanation of the anisotropic spectral
density~\cite{oo_nematic_2011}. Finally, as a third way to induce an
anisotropy, we make hoppings along one lattice direction five to ten
percent larger. 

Following~\cite{Akw_nematic}, this Hamiltonian  is treated with
the variational cluster approach. This method allows to include
correlations within a small cluster (4 sites for the
three-orbital model discussed here), which is solved almost exactly with
Lanczos exact diagonalization. Hopping between the clusters is then 
included as a
perturbation~\cite{Gros:1993p2667,Senechal:2000p2666}. Long-range
order can be treated  by optimizing the grand
potential with respect to fictitious ordering
fields~\cite{Aic03,Dahnken:2004p2409}, the SDW state of a two-band
model for pnictides has been studied with this
approach~\cite{Daghofer:2008,Yu:2009p2127}. In fact, all 
parameters of the one-particle part of the Hamiltonian can in principle
be optimized, we found here that optimizing an overall fictitious
chemical potential is necessary near the SDW transition to obtain a
stable solution.

\section{Results}\label{sec:results}

\begin{figure}
\centering
\includegraphics[width=0.47\textwidth]{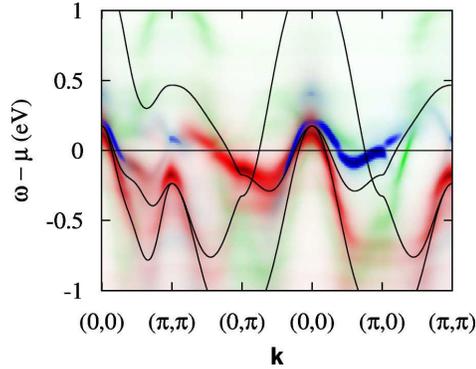}
\caption{Spectral density $A({\bf k},\omega)$ for onsite interactions
  near the SDW transition ($U=1.02\;\textrm{eV}$, $J=U/4$) and a
  phenomenological field $\Delta=0.1\;\textrm{eV}$, see
  (\ref{eq:oo}), breaking the orbital $xz$/$yz$ symmetry and
  favouring the $xz$ orbital. Solid lines give the noninteracting
  bands in terms of the pseudo-crystal momentum ${\bf
    \tilde{k}}$; shading gives the spectral weight of
    the interacting system for `lab-space' momentum ${\bf k}$, $xy$
    weight is thus shifted by $(\pi,\pi)$ with respect to ${\bf
    \tilde{k}}$. In
  the online colour figure, red, blue and green shading illustrate
  spectral weight in the $xz$, $yz$, and $xy$ orbitals respectively.\label{fig:Akw_oo}}
\end{figure}

\begin{figure}
\centering
\includegraphics[width=0.47\textwidth]{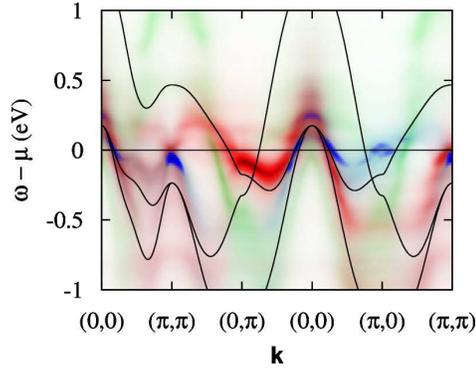}
\caption{As figure~\ref{fig:Akw_oo}, but without orbital energy
  splitting. Instead, there are anisotropic AFM couplings, see
  (\ref{eq:Heisenberg}), acting
  within the directly solved four-site cluster, i.e., on a very short
  distance only. $J_x = 0.04\;\textrm{eV} \gg J_y = 0.01\;\textrm{eV}$ \label{fig:Akw_AF_AF}} 
\end{figure}

It has been shown that both orbital ordering~\cite{oo_nematic_2011} and anisotropic magnetic
correlations~\cite{Akw_nematic} can reproduce the band distortions in
a manner broadly consistent with ARPES, i.e., the $yz$ states at
$X=(\pi,0)$ move to higher energies than the $xz$ states at
$Y=(0,\pi)$. Differences between these two ways of breaking rotational
lattice symmetry mostly affect states around $\Gamma=(0,0)$, where an
orbital energy splitting induces stronger distortions than the
magnetic fluctuations in the absence of onsite
interactions~\cite{Akw_nematic}. As can be seen by comparing
figures~\ref{fig:Akw_oo} and~\ref{fig:Akw_AF_AF}, this remains
true when onsite interactions push the system closer to the SDW
transition. One may also note that despite the sizable onsite
interactions, one still needs an orbital energy splitting 
$\Delta=0.1\;\textrm{eV}$ to induce the band anisotropy between $X$
and $Y$ as seen
in figure~\ref{fig:Akw_oo}. The anisotropy in the final band structure is here not even extreme,
with 
the $yz$ states not quite reaching the Fermi level, and $\Delta$ of a
similar order of magnitude induces comparable distortions in non-interacting
bands~\cite{oo_nematic_2011,Akw_nematic}. This is in contrast to short-range magnetic couplings that
become more effective at distorting the bands when onsite interactions
are switched on~\cite{Akw_nematic}. Onsite interactions thus enhance
the tendency to short-range magnetic correlations but do not 
appear to strengthen tendencies to a direct orbital energy splitting
in a comparable manner. We did not find spontaneous symmetry breaking
  between the $xz$ and $yz$ orbitals, which can be investigated in the
  variational cluster-perturbation theory~\cite{Aic03,Dahnken:2004p2409}.

\begin{figure}
\centering
\includegraphics[width=0.47\textwidth]{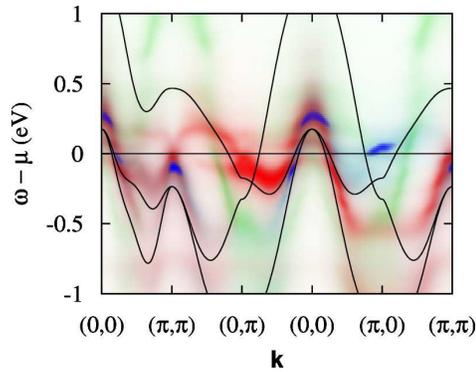}
\caption{As figure~\ref{fig:Akw_oo}, but without orbital energy
  splitting. Instead, there is a
  phenomenological difference in hopping parameter $t_2$: it is
  $10\;\%$ larger along $y$-direction. \label{fig:Akw_t2}} 
\end{figure}

Figure~\ref{fig:Akw_t2} illustrates an alternative way to induce an
anisotropy: The hopping parameter $t_2$ is chosen $10\;\%$ larger along
the $y$-direction, a rather larger hopping anisotropy. For non-interacting
electrons (not shown), this simply increases the dispersion somewhat, but does not
raise the states at $Y$. As can be seen in figure~\ref{fig:Akw_t2}, the
interplay of the hopping anisotropy with onsite interactions, which favour a $(\pi,0)$ or $(0,\pi)$
SDW, distorts the bands: The
$yz$ band going from $\Gamma$ to $X$ has become very incoherent, and
the only remaining coherent states at $X$ are above the Fermi
level. It should be noted that the effect of an 
anisotropy on various hopping parameters is not consistent: It is
largest for $t_2$, which is the larger NN hopping entering the kinetic
energy of the $xz$ and $yz$ orbitals, see (\ref{eq:ekin}). $10\;\%$
anisotropy of the other NN hoppings $t_1$, $t_5$ and $t_7$ leads to far
smaller effects (not shown).

Finally, we take a closer look at the short-range magnetic
interactions that were shown to induce band anisotropies
in~\cite{Akw_nematic}, where the strength of the AFM 
(along $x$) and FM (along $y$) couplings were chosen 
with opposite sign, but of equal strength. By varying the two
parameters independently, we found that it is not necessary to have
one FM and one AFM direction. For example, figure~\ref{fig:Akw_AF_AF}
shows a spectrum obtained for a case where both couplings are AFM, but
the one along $x$ is much stronger ($J_x=0.04\;\textrm{eV}$) than the
one along $y$ ($J_y=0.01\;\textrm{eV}$). The AFM couplings act only 
within the directly solved cluster, i.e., they favour AFM bonds more
along $x$ than along $y$. The clusters are coupled within
cluster-perturbation theory only via the kinetic energy, i.e., there
is no long-range magnetic order. The anisotropic signatures seen in
the spectral density are comparable to those found for
$J_x=-J_y=0.015\;\textrm{eV}$~\cite{Akw_nematic}. As
  the order parameter for nematic order is of higher order, its
  spontaneous symmetry breaking can not be studied in
  cluster-perturbation theory.

\section{Discussion and Conclusions}\label{sec:conclusions}

We investigated how various mechanisms of breaking the four-fold
lattice symmetry of a three-orbital model for Fe-As planes manifest
themselves in the spectral density. Onsite interactions bring the
system here close to a SDW transition, where
short-range magnetic couplings have previously been found to be more
effective at breaking rotational symmetry than in the non-interacting
system. We find that an orbital energy splitting and anisotropic
hoppings can lead to qualitatively similar features in the interacting
bands as short-range magnetic correlations: the states at $X$ can be
found at higher energies than those at $Y$ and can even move up to the
Fermi level, see also previous
studies~\cite{oo_nematic_2011,Akw_nematic}. The band/orbital
anisotropy near the Fermi surface can be similarly pronounced both in cases with
a strong total orbital polarization $n_{xz}-n_{yz}\approx 0.38$ (found
for $\Delta =0.1\;\textrm{eV}$ as in figure~\ref{fig:Akw_oo}) and with
nearly vanishing polarization $n_{xz}-n_{yz}\approx 0.02$ (found
for the anisotropic AFM couplings as in
figure~\ref{fig:Akw_AF_AF}). A splitting of $\approx
  100\;\textrm{meV}$ between the $xz$ at $Y$ and the $yz$ states at
  $X$, which moves the latter close to or just above the chemical
  potential, can be induced by (i) an orbital energy splitting of
  $\Delta 
  =0.1\;\textrm{eV}$, (ii) $10\;\%$ anisotropy in the hopping
  parameter $t_2$ and (iii) a magnetic anisotropy of $30\;\textrm{meV}$
between the $x$ and $y$ directions.

There are differences in the results obtained in the three scenarios: In the case of an
orbital splitting, the changes around $\Gamma=(0,0)$ are more
pronounced than in either the magnetic scenario of the scenario with
anisotropic hopping. In the latter case, the spectra appear less
coherent even near the Fermi level, and the $yz$ band going from
$\Gamma$ to $X$ almost disappears except for states very close to $X$
that are above the Fermi level. Comparing to ARPES~\cite{Yi:PNAS2011,ARPES_NaFeAs11,He:2010pNaFeAs,ZhangARPES_NaFeAs_2011}, this lack of
coherence does not appear to be in good agreement. Distinction between
the other two scenarios is more difficult and has previously been
discussed for an orbital energy difference in the non-interacting
model~\cite{Akw_nematic}. The conclusions drawn there remain valid for
the interacting model: ARPES data look more consistent with slighter
changes around $\Gamma$ than those arising from an orbital energy
difference large enough to raise the states at $X$ to the Fermi level. However, it has to be stressed that
both our model (including only the three most important orbitals) and
our method (where the impact of onsite correlations is only treated
exactly within a very small four-site cluster) imply substantial
approximations. The most robust conclusion to be drawn might thus be
that a decision between the scenarios based on experimental ARPES data
remains difficult. Finally, it is also thinkable that different
mechanisms are driving the breaking of the fourfold lattice symmetry
in various compounds of the pnictide family.

\ack 
This research was supported by the DFG 
under the Emmy-Noether programme and via the Graduate Training Group
GRK 1621. 

\section*{References}


\end{document}